\documentclass[prd,10pt,amsmath,amssymb,aps,nofootinbib,showkeys,superscriptaddress,showpacs,notitlepage]{revtex4-1}
\usepackage{graphicx, epsfig, amssymb} %include figure files
\usepackage{amsmath, amsfonts}
\usepackage{bm} %include bold math: \bm{} creates bold letters in math mode
\usepackage{color}
\usepackage[usenames]{xcolor}
\usepackage{hyperref}
\usepackage{siunitx}
\hypersetup{linktocpage,colorlinks=true,urlcolor=blue,linkcolor=blue,citecolor=blue}  
\usepackage{verbatim}
\usepackage[utf8]{inputenc}
\usepackage{natbib}

\newcommand{\rd}{\mathrm{d}} % Roman d for differential
\newcommand{\ri}{\mathrm{i}} % Roman i for imaginary number
\newcommand{\tr}{\mathrm{tr}}
\newcommand{\rD}{\mathrm{D}}
\newcommand{\rt}{\mathrm{t}}
\newcommand{\ta}{\tilde{\alpha}} 

\begin{document}
 
\title{Meronic AdS Black Holes in Gauss--Bonnet Theory}
\author{Daniel \surname{Flores-Alfonso}}
\email[]{daniel.flores@xanum.uam.mx}
\affiliation{Departamento de F\'isica,
Universidad Aut\'onoma Metropolitana - Iztapalapa,\\
Av. San Rafael Atlixco 186, C.P. 09340, Ciudad de M\'exico, Mexico}

\author{Bryan O. \surname{Larios}}
\email[]{bryan.larios@unah.edu.hn}
\affiliation{Escuela de F\'isica, Facultad de Ciencias, 
Universidad Nacional Aut\'onoma de Honduras,\\ 
Blvr. Suyapa, Tegucigalpa, Municipio del Distrito Central 11101, Honduras}
\affiliation{Mesoamerican Centre for Theoretical Physics, 
Universidad Aut\'onoma de Chiapas,\\ 
Carretera Zapata Km. 4, Real del Bosque (Ter\'an), 29040, Tuxtla Guti\'errez, Chiapas, Mexico}


\begin{abstract}
We examine analytical, intrinsically non-Abelian, black holes with SU(2) Yang--Mills matter content.
Working in the extended thermodynamics scenario of Lovelock
black holes we study the phase structure of a four and a five-dimensional spherically symmetric configuration. We work in Gauss--Bonnet--Einstein--Yang--Mills gravity with negative cosmological constant and use 
Euclidean methods to explore the thermodynamics of the systems. We observe that the solutions belong to the expected universality class of van der Waals and find a reentrant phase transition.
\end{abstract}

\maketitle

\section{Introduction and Motivation}

Taking into consideration both quantum and spacetime physics has shown that black holes are indeed thermodynamic systems.
Black holes radiate heat in a manner quite similar to black body radiation.
It is now a famous result that, the temperature of a black hole is given by its surface gravity and that
the horizon area corresponds to the entropy~\cite{Bekenstein:1973ur,Bekenstein:1974ax,Hawking:1974sw,Hawking:1976de}.
The archetypal example is the Schwarzschild black hole, which is completely determined by its mass $M$.
This single parameter family is characterized by a temperature of $T=1/8\pi GM$ and an entropy of $S=4\pi G M^2$.
The relationships between the gravitational charge $M$ and the thermodynamic quantities $T$ and $S$ yield
\begin{equation}
 M=2TS,\quad \text{and} \quad \rd M=T\rd S. \label{smarr4}
\end{equation}
These algebraic and differential relations establish the mass as the thermodynamic internal energy
of the system.
The first equation in \eqref{smarr4} is Smarr's mass formula while the second is the first law of black hole thermodynamics~\cite{Smarr:1972kt,Bardeen:1973gs}. 

From the point of view of thermodynamics, Smarr's relation is a type of Gibbs--Duhem equation.
The factor of two appearing in the formula is indicative that black hole thermodynamics is not exactly like that of ordinary substances such as
fluid mixtures, ferromagnetic dipoles or quantum gases. Common substances are characterized by 
homogeneous fundamental relations, e.g., $S(\lambda U,\lambda V)=\lambda S(U,V)$.
The Schwarzschild black hole is a homogeneous thermodynamic system of degree two, i.e., $S(\lambda M)=\lambda^2S(M)$.
When it exists, the degree of homogeneity of a fundamental relation can be read off from its associated Gibbs--Duhem equation.
This highlights the significance of a black hole's Smarr formula, see for example~\cite{Bravetti:2017peq}.

In this manuscript, we are interested in spherically symmetric spacetimes.
Birkhoff's theorem asserts that any spherically symmetric vacuum solutions of General Relativity (GR)
must be static and asymptotically flat. In other words, the unique solution in four dimensions is the Schwarzschild metric.
In dimensions higher than four, the equivalent solutions were derived by Tangherlini~\cite{Tangherlini:1963bw}.
For any given dimension, $D$ say, the first law of black hole thermodynamics is as in \eqref{smarr4}. Regardless, the Smarr
relation generalizes in accordance with two basic ideas, (i) the action of gravity when higher spatial dimensions 
are available and (ii) the holographic principle. In general, the mass formula takes the form
\begin{equation}
 (D-3)M=(D-2)TS. \label{smarrD}
\end{equation}

Until now, the equations we have written down lack the typical pressure-volume terms found in the first law of
thermodynamics and the Gibbs--Duhem equation (cf. ~\cite{Dolan:2012jh}). 
With this in mind, let us recall that the cosmological constant $\Lambda$ admits a perfect fluid interpretation, 
where the isotropic pressure is given by $p=-\Lambda/8\pi G$. One way to consider this pressure 
as actually thermodynamical arises when $\Lambda$ is not a universal constant but rather a constant of 
motion~\cite{Henneaux:1984ji,Teitelboim:1985dp,Henneaux:1985tv,Henneaux:1989zc}. Furthermore, the Euclidean approach to quantum gravity~\cite{Gibbons:1976ue,Hawking:1982dh} is consistent with this idea, as its methods are on-shell. This method relates the Euclidean on-shell action to the
free energy of the system through a bridge equation~\cite{Dolan:2010ha}. Thus, considering $\Lambda$ as a thermodynamic variable that remains fixed in the action is
comparable to isobaric ensembles\cite{Dolan:2013ft}. As a consequence, variation of the free energy entails a variation with respect to the thermodynamic pressure.

Crucially, note that the original Smarr relation for Kerr black holes is modified in the presence of a non-zero 
cosmological constant~\cite{Caldarelli:1999xj}. Thus, adopting $\Lambda$ as a thermodynamical variable
endows meaning to the Smarr relation of Kerr--Newman-AdS black holes\cite{Hyun:2017nkb}.
The extension of the classic framework of black hole thermodynamics leads to a new interpretation of black hole mass.
The extended first law of thermodynamics and Gibbs--Duhem relation determine the mass to be the enthalpy of the system rather than its internal energy~\cite{Kastor:2009wy,Dolan:2011xt}.
The thermodynamic conjugate of pressure $p$ is, of course, referred to as \emph{volume}. However, this thermodynamic variable, $V$ say, need not be
related to any geometrical volume\cite{Cvetic:2010jb,Johnson:2014xza}. This extended thermodynamics can be further expanded when considering Lovelock gravity.
This is to say, all Lovelock couplings
are considered as thermodynamic variables~\cite{Kastor:2010gq}.
Our focus here is Gauss--Bonnet (GB) theory, meaning both the cosmological constant $\Lambda$ and the GB coupling $\alpha$ are considered as thermodynamic variables.
It should be understood, for consistency, that these thermodynamic variables are held fixed in the action. 

Let us assume for the time being that the pressure and volume variables are not in play.
The spherically symmetric Gauss--Bonnet vacua is the Boulware--Deser black hole~\cite{Boulware:1985wk}.
For a zero-valued cosmological constant the black hole family is asymptotically flat. Thus, aside from the black hole mass the only additional thermodynamic variable is the GB parameter $\alpha$. Since both the mass and the entropy depend on the horizon geometry this entails a modification of
the first law of thermodynamics within the extended framework, i.e.,
\begin{equation}
 \rd M=T\rd S -\psi\rd\alpha.
\end{equation}
Just above $\psi$ is the thermodynamic conjugate of $\alpha$, see Appendix \ref{secc:Boulware--Deser} for further details. Smarr's approach leads to an algebraic equation
which is exactly of Gibbs--Duhem type
\begin{equation}
 (D-3)M=(D-2)TS-2\psi\alpha. \label{smarrBD}
\end{equation}
Extended thermodynamics provides the necessary conceptual structure for this Smarr relation to admit
a thermodynamical interpretation. Moreover, the Gauss--Bonnet parameter has units of length squared, this is consistent with
the scaling factor of 2 in \eqref{smarrBD}. Entropy is given by horizon area (plus corrections due to $\alpha$)
consistent with the factor $D-2$ in the equations above. As a thermodynamic variable $\alpha$ is very closely related to entropy,
which is why its dual, $\psi$, is closely related to temperature [cf. Appendix \ref{secc:Boulware--Deser}].

Of course, various spherically symmetric spacetimes have been studied beyond vacuum Gauss--Bonnet. Lovelock gravity specially, where many static solutions have been investigated~\cite{Dotti:2006cp,Kastor:2006vw,Cai:2006pq,Dotti:2007az,Garraffo:2008hu,Dadhich:2012ma}.
Therein, black hole solutions have been generalized in such a way that their spherical horizon has been substituted by to some other appropriate manifold.
In addition, solutions may be generalized to electrovacuum, as in
the classic thermodynamic framework of GR.
It is well-known that electromagnetic fields can also be included into the Lovelock scenario. Matter fields from Maxwell theory and nonlinear electrodynamics have been studied~\cite{Wiltshire:1985us,Banados:1993ur,Aiello:2004rz,Aiello:2005ef,Dehghani:2008qr} generalizing solutions
in GR~\cite{Demianski:1986wx,Cai:2004eh,Dey:2004yt}. 
Many of the previous scenarios have been investigated in the context of extended thermodynamics as well (see \cite{Kubiznak:2016qmn} for a recent review).
Some examples include references~\cite{Gunasekaran:2012dq,Gonzalez:2015mpa} where the focus is
nonlinear constitutive relations.

Maxwell theory is a special case of Yang--Mills (YM) theory when the gauge group is chosen as U(1).
We know from the standard model of particles that YM fields represent interactions beyond
the electromagnetic. Electroweak theory is an example that together with Einstein--Maxwell gravity
point naturally towards the study of Einstein--Yang--Mills equations. For SU(2) particle-like and
black hole solutions have been known for some time~\cite{Bartnik:1988am,Volkov:1989fi,Kuenzle:1990is,Bizon:1990sr}; for a review
on the subject we refer to~\cite{Volkov:1998cc}. It should be mentioned that the first results in this area were numerical and many efforts have been made since towards finding and understanding analytical solutions. A relatively recent attempt to move forward in this direction is to construct self-gravitating meron configurations~\cite{Canfora:2012ap,Canfora:2018ppu}.
Merons were originally proposed in~\cite{deAlfaro:1976qet}, they are essentially non-Abelian and arguably simple in nature.

As mentioned above, genuinely non-Abelian self-gravitating YM configurations are desirable.
Very few exact solutions of this nature are known in the literature.
We stress on the non-Abelian quality of the field content as it is often unclear whether a given
configuration, in fact, belongs to an Abelian sector of the theory. Furtherstill, as opposed to the Maxwell case, there is no uniqueness theorem for YM black holes. However, the general thermodynamics
of YM black holes has been carried out (in the classic setting) in reference~\cite{Heusler:1993cj}.
Therein, the YM configuration is analyzed alongside another nonlinear matter model, that of Skyrme.
This model is very useful in particle and nuclear physics as it is closely related to
the low energy limit of QCD~\cite{Witten:1983tx}. 
Additionally, we mention that Einstein--Skyrme systems~\cite{Canfora:2013osa} have been recently analyzed in~\cite{Flores-Alfonso:2019aae} which point to a close relationship with charged AdS black holes. 

Our objective is to study meronic black holes in Gauss--Bonnet theory. We
focus on generalizing known Einstein solutions and examining the effects of the GB parameter.
We present a four dimensional Reissner--Nordstr\"om-like black hole with entropy modified by $\alpha$. This solution is closely related to those studied in~\cite{Canfora:2012ap,Castro:2013pqa,Xu:2019xif}.
Note that in four dimensions the GB contribution is a boundary term that does not affect the dynamics but makes a difference in the calculation of conserved quantities~\cite{Aros:1999id}.
We also present a five dimensional black hole with a Boulware--Deser-esque metric function.
The structure of the solution generalizes black holes such as those found in references~\cite{Okuyama:2002mh,Canfora:2018ppu}.

This paper is organized as follows: In Section \ref{secc:AAAA} we present the Einstein--Gauss--Bonnet gravity theory minimally coupled to Yang--Mills fields. We provide the equations of motion and detail the field and symmetry assumptions we use throughout the manuscript. We end the section describing the Euclidean approach undertaken in subsequent sections. In Section \ref{secc:4D} we inject a self-gravitating four-dimensional Einstein meron into Gauss--Bonnet theory and explore the consequences of the theory's parameter $\alpha$. The dynamics the meron obeys is Einstein--Yang--Mills as the Gauss--Bonnet term is topological in this dimension, however it does affect the thermodynamics.
We study the configuration's thermodynamics and phase structure.
In Section \ref{secc:5D} we generalize a five dimensional Einstein--Yang--Mills solution to the Lovelock scenario. After detailing the singularity structure of the new black hole we establish which known solutions it interpolates. We interpret the parameters of the solution in accordance with its thermodynamics. We compute the thermodynamic state equations and describe the solution's phase transitions. We end the section with comments about certain special values of the Lovelock coupling constants. Lastly, in Section \ref{secc:Conc} we write the concluding remarks of our work.

In Appendix \ref{secc:Boulware--Deser} we provide a way to compute $\psi$, the thermodynamic conjugate of $\alpha$,
when mass and entropy are known as functions of the horizon radius $r_+$ and the GB parameter $\alpha$. The definition of horizon radius
and its differential yield functions for the mass and the temperature (via the Hawking prescription). This in turn gives the corresponding entropy function, see e.g., reference~\cite{Johnson:2015ekr}.

\section{Action, Ans\"atze and Anti-de Sitter} \label{secc:AAAA}

General Relativity and Gauss--Bonnet gravity are special cases of Lanczos--Lovelock theory~\cite{Lanczos:1938sf,Lovelock:1971yv}.
Lovelock's theorem establishes GR as the most general metric theory of gravity, in four-dimensional vacuum,
which has symmetric, divergence free and second order equations of motion\footnote{The fundamental assumption that the equations of motion be second order is justified as it systematically prevents the appearance of Ostrogradsky instabilities.}.
Lovelock constructed, exhaustively, all second rank
tensors which comply with these properties for any arbitrary dimension. 
Lovelock gravity is ghost free~\cite{Zumino:1985dp}
and as far as propagation is concerned, the theory has the same degrees of freedom as GR for any given dimension~\cite{Henneaux:1990au}.
In this paper, we are only concerned with dimensions four and five for which the Lanczos--Lovelock action corresponds to that of Gauss--Bonnet. It should be noted that in dimension four the Gauss--Bonnet contribution is non-dynamical and so the equations of motion are Einstein's.
As a final introductory remark we mention that GB theory appears in string theory as a low-energy effective action~\cite{Zwiebach:1985uq}. In other words, the Gauss--Bonnet term corrects
the stringy field equations in dimensions higher than four.

Let us write down the Gauss--Bonnet gravity action minimally coupled to SU(2) Yang--Mills matter
\begin{equation} \label{LLaction}
 I[g,A]=\frac{1}{16\pi G}\int \rd^Dx\sqrt{-g}\left(R-2\Lambda+\alpha{\cal L}_{\rm GB}\right) 
 +\frac{1}{8\pi e^2}\int \rd^Dx\sqrt{-g}~\tr\langle F,F\rangle,
\end{equation}
with $G$ and $e$ Newton's constant and the YM coupling in spacetime dimension $D$, respectively.
In the previous equation we have used
\begin{equation}
 {\cal L}_{\rm GB}=R^2-4R_{ab}R^{ab}+R_{cdab}R^{cdab},
\end{equation}
as shorthand for the Gauss--Bonnet contribution to the gravity functional
and $F$ for the YM field strength
\begin{equation}
 F=\rd A+\frac{1}{2}[A,A]. \label{defF}
\end{equation}
We have also denoted by $\langle\cdot,\cdot\rangle$ the inner product of differential forms.
Let's recall that $F$ is locally represented by an $\mathfrak{su}$(2)-valued two-form.
We conventionally use $\rt^1,\rt^2$ and $\rt^3$ to represent the linear generators of $\mathfrak{su}$(2) as well as the unit
vectors in $\mathbb{C}^2$, which model SU(2). These matrices are traceless and have Frobenius norm $1/\sqrt{2}$. They also comply with the commutation relations
\begin{equation}
 [\rt^i,\rt^j]=\ri\epsilon_{ijk}\rt^k.
\end{equation}

Varying the action functional \eqref{LLaction} with respect to the metric $g$ yields
\begin{equation}
 R_{ab}-\frac{1}{2}Rg_{ab}+\Lambda g_{ab}+\alpha H_{ab}=8\pi G\, T_{ab}, \label{EOM}
\end{equation}
where we have defined a tensor $H$ which has components
\begin{equation}
  H_{ab}=2RR_{ab}-4R_{ac}R^{c}_{~~b}-4R^{cd}R_{acbd}
+2R_{a}^{~~cde}R_{bcde}-\frac{1}{2}{\cal L}_{GB}\, g_{ab}.
\end{equation}
In equation \eqref{EOM} the energy-momentum tensor $T$ is given by
\begin{equation}
 T_{ab}=-\frac{1}{4\pi e^2}\tr\left(F_{ac}F_{b}^{~c}-\frac{1}{2}g_{ab}\langle F,F\rangle\right).
\end{equation}
Notice that for two forms $\langle F,F\rangle=F_{ab}F^{ab}/2$.
When varying the action with respect to the gauge potential $A$ the resulting equation is
\begin{equation}
 \rD\star F=0, \label{YMeom}
\end{equation}
where $\star$ is the Hodge star linear map and $\rD=\rd+[A,~]$ is the covariant derivative --- in the YM sense.

\subsection{Meron and Symmetry Ans\"atze} \label{secc:masa}
In order to produce intrinsically non-Abelian self-gravitating configurations, we opt for a meron Ansatz, i.e.,
\begin{equation}
 A=\lambda\omega\quad (\lambda\neq 0,1), \label{meronansatz}
\end{equation}
where $\lambda$ is a constant and $\omega$ satisfies the following equation
\begin{equation}
 \rd\omega+\frac{1}{2}[\omega, \omega]=0.\label{meronEq}
\end{equation}
Thence, using \eqref{defF} and the previous equations we see straightforwardly that
\begin{align}
 F&=\lambda\rd \omega +\frac{1}{2}\lambda^2[\omega, \omega],\notag\\
 &=-\lambda\frac{1}{2}[\omega, \omega]+\frac{1}{2}\lambda^2[\omega, \omega],\notag\\
 &=\frac{1}{2}\lambda(\lambda-1)[\omega, \omega]\label{meron}.
\end{align}
In other words, the trivial nature of $\omega$ \eqref{meronEq} guarantees that $A$ is non-trivial.

The self-gravitating Yang-Mills fields that we consider in the upcoming Sections are purely magnetic.
It is known that under this circumstance the supporting spacetime must be static~\cite{Heusler:1993cj}. While it was proved in~\cite{Smoller:1995nk} that static configurations of this type
(that are well-behaved in the far field) are either black holes, particle-like solutions or of Riessner--Nordstr\"om‐like nature.
Additionally, the merons we investigate produce spherically symmetric energy-momentum tensors. In other words, the background they generate are spherically symmetric as well.

In general, a spherically symmetric spacetime admits Schwarzschild coordinates such that the metric is written as
\begin{equation}
 g=-a(r)\rd t\otimes \rd t+b(r)\rd r\otimes \rd r+r^2\gamma_{D-2},
\end{equation}
with $\gamma$ the round metric on the nested (hyper)spheres.
Further simplification is possible once the equations of motion are taken under consideration, e.g., the Einstein merons of \cite{Canfora:2012ap,Canfora:2018ppu} have geometries of the form
\begin{equation}
 g=-f(r)\rd t\otimes \rd t+f(r)^{-1}\rd r\otimes \rd r+r^2\gamma_{D-2}. \label{SchCoords}
\end{equation}
This symmetry, in turn, implies that the Lovelock field equations reduce to a single ordinary differential equation for the metric
function $f$. This equation integrates to produce an algebraic equation for $f$\cite{Wheeler:1985nh,Wheeler:1985qd}. This polynomial has degree $n$, given by the
highest-order curvature term in the Lanczos--Lovelock action which contributes to the equations of motion, i.e., $n=[(D-1)/2]$,
where closed brackets indicate taking the integer part.
The most compact way to write these Wheeler polynomials is by introducing and auxiliary function ${\cal F}=(1-f)/r^2$.
A Wheeler polynomial $P({\cal F})$ satisfies the following equation
\begin{equation}
 P({\cal F})\equiv\sum\limits^{n}_{i=0} a_i {\cal F}(r)^i=\frac{s}{r^{D-1}}+S(r), \label{Wheeler}
\end{equation}
where the $a_i$ coefficients are determined by the Lovelock couplings and
$s$ is an integration constant. For many black holes, the constant $s$ corresponds to the mass, e.g., Schwarzschild--Tangherlini. The source function $S$ is 
derived from the energy-momentum tensor, of course, for vacuum solutions $S(r)=0$.
For recent applications we refer the reader to~\cite{Oliva:2010eb,Giribet:2014bva,Ray:2015ava,Lagos:2017vap,Chernicoff:2016qrc,Corral:2019leh}. In what follows we use Wheeler polynomials to concisely
present the meronic black holes under consideration.

\subsection{Anti-de Sitter Boundary Counterterms} \label{secc:ABC}
In this paper, we focus on solutions to equations \eqref{EOM} and \eqref{YMeom} with a negative cosmological constant, in order to describe asymptotically Anti-de Sitter (AdS) spacetimes.
The defining length scale $l$ of AdS is given by its radius of curvature which is related to the cosmological constant by
\begin{equation}
\Lambda=-\frac{(D-1)(D-2)}{2l^2}. \label{Lambda} 
\end{equation}
Analogously, the GB coupling constant is often rewritten as
\begin{equation}
 \alpha=\frac{\ta}{(D-3)(D-4)}. \label{alpha}
\end{equation}
This way of parameterizing the Lovelock coupling constants appears naturally in
Wheeler polynomials, fixing the constants $a_i$ in equation \eqref{Wheeler}. 

To analyze the thermodynamics of the aforementioned black holes we use Euclidean methods.
Thus, consider the Euclidean version of action \eqref{LLaction}
\begin{equation} \label{Eaction}
 I_B=-\frac{1}{16\pi G}\int \rd^Dx\sqrt{g}\left(R-2\Lambda+\alpha{\cal L}_{\rm GB}\right) 
 -\frac{1}{8\pi e^2}\int \rd^Dx\sqrt{g}~\tr\langle F,F\rangle,
\end{equation}
where we have adhered to the conventions of reference~\cite{Emparan:1999pm}.
As fas as the gravitational sector is concerned, if the boundary metric is fixed beforehand then variation with respect to the bulk metric will not yield \eqref{EOM} unless the functional is supplemented by surface terms at the boundary.
For the Einstein part of the equations of motion the boundary term is known as the Gibbons--Hawking
term~\cite{Gibbons:1976ue}. For the Gauss--Bonnet component the term was given by Myers in~\cite{Myers:1987yn}. Here, we use the notation of reference~\cite{Davis:2002gn} to write both surface integrals as
\begin{equation}
I_{ST}=-\frac{1}{8\pi G}\int \rd^{D-1}x\sqrt{h}\left[K+2\alpha(J-2\widehat{G}^{ab}K_{ab})\right], \label{GHM}
\end{equation}
where $K_{ab}$ are the components of the extrinsic curvature tensor of the boundary and $K$
designates its trace. The induced metric on the boundary is $h$ and $\widehat{G}$ represents its 
Einstein tensor. Finally, $J$ is the trace of a tensor defined by
\begin{equation}
 J_{ab}=\frac{1}{3}\left(2K K_{ac}K^c_{~b}+K_{cd}K^{cd}K_{ab}-2K_{ac}K^{cd}K_{db}-K^2 K_{ab}\right).
\end{equation}
Notice that the surface integrals in \eqref{GHM} depend on the bulk metric in the sense that the integrands depend on extrinsic curvatures. We point this out because there is a crucial ambiguity
which arise from the following fact. An arbitrary surface integral over the boundary may be added
as long as in only depends on the intrinsic curvature of the boundary. This includes integrands
that only depend on the fixed metric $h$ and its scalar curvature $\widehat{R}$. For AdS this ambiguity has been resolved for some time now, at least up to dimensions relevant for string theory.

This counterterm method originated with the AdS/CFT correspondence but the main purpose it serves here is that it allows for the calculation of a finite Euclidean action. For an in depth description
of how this method came to be we refer the reader to~\cite{Emparan:1999pm} and references therin.
The counterterm integral is arrangeable as a power series over the boundary's intrinsic curvature and its derivatives. We truncate this power expansion up to orders relevant for dimensions four and five, i.e.,
\begin{equation}
 I_{CT}=\frac{1}{8\pi G}\int \rd^{D-1}x\sqrt{h}\left(\frac{D-2}{l}
 +\frac{l}{2(D-3)}\widehat{R}\right).
\end{equation}

Now, concerning the matter action, not all actions require additional counterterms, e.g., Maxwell
in spacetime dimension four. However, surface integrals as counterterms for specific matter content have been developed. Close examples to YM merons are skyrmions and axions, for which counterterms have been successfully applied~\cite{Flores-Alfonso:2019aae,Caldarelli:2016nni}.

\section{The Four-Dimensional Meron} \label{secc:4D}

A distinctive aspect of non-Abelian theory is that 
a given field strength does not determine a unique gauge potential
up to gauge transformations.
This profound attribute was first noticed in reference~\cite{Wu:1975vq}.
One may, in fact, construct many (gauge inequivalent) gauge potentials which
yield the same field strength. Albeit, the field content is distinguishable through
higher curvature invariants. Merons provide an easy way for constructing such examples~\cite{Deser:1976wj}. For the self-gravitating kind of merons an example is provided by an Abelian Reissner--Nordstr\"om-like solution~\cite{Smoller:1997qr} and a meronic black hole which has Reissner--Nordstr\"om (RN) geometry~\cite{Canfora:2012ap}. 

The meron we analyze in this Section has a gauge field of the following type
\begin{equation}
 A=\lambda U^{-1}\rd U. \label{omegaU}
\end{equation}
where $U$ is an SU(2)-valued scalar field, in the adjoint representation.
The field is such that it yields a
spherically symmetric background. We choose to coordinate the metric as in \eqref{SchCoords}
\begin{equation}
 g=-f(r)\rd t\otimes \rd t+f(r)^{-1}\rd r\otimes \rd r+r^2\gamma_{2},
\end{equation}
with $\gamma_2$ charted by spherical angles $(\vartheta,\varphi)$.
In these coordinates the scalar field is
\begin{equation}
U=\cos\vartheta \rt^3+\sin\vartheta\sin\varphi \rt^2+\sin\vartheta\cos\varphi \rt^1. \label{U}
\end{equation}
Notice that the value of $U$ is traceless and has Frobenius norm $1/\sqrt{2}$, placing it on SU(2). The origin of this scalar field is a hedgehog Ansatz~\cite{Canfora:2012ap}. 
From \eqref{U} it can be verified that the field strength is purely magnetic
and that the energy-momentum tensor is spherically symmetric.
In fact, the latter coincides (up to a scaling) with the energy-momentum tensor of 
the famous Dirac monopole~\cite{Dirac:1931kp}.

The YM equations of motion only hold if $\lambda=1/2$, in correspondence
with the seminal work of~\cite{deAlfaro:1976qet}.
The gravity equations are Einstein and lead to the metric function being of RN type
\begin{equation}
 f(r)=1-\frac{2Gm}{r}+\frac{\rho^2}{r^2}+\frac{r^2}{l^2},
\end{equation}
consistent with the previous paragraph. The equations of motion also determine $\rho$ to be
\begin{equation}
 \rho^2=\frac{G}{2e^2}. \label{rhocc}
\end{equation}
Notice that this quantity is not an integration constant, it is fixed by the couplings 
of the theory. However, many parallels exist between $\rho$ and the electromagnetic charge
of an RN black hole. For example, consider the energy of the meron
\begin{equation} \label{YMenergy}
 E\equiv-\frac{1}{8\pi e^2}\int \tr\langle F,F\rangle \star k=\frac{\rho^2}{2Gr_+},
\end{equation}
where $k$ is the timelike Killing form and $r_+$ is the horizon radius. Here again $\rho$ plays the role the Abelian charge does
in the RN solution.

As a final introductory remark and for illustrative purposes concerning the next Section we write down the Wheeler polynomial \eqref{Wheeler} for this meron configuration
\begin{equation}
 \frac{1}{l^2}+{\cal F}+\ta{\cal F}^2=\frac{2Gm}{r^3}-\frac{\rho^2}{r^4}.
\end{equation}
For this dimension $\ta=0$, so what would be a quadratic equation for Gauss--Bonnet theory
is instead linear as, once more, the dynamics are Einstein. In the next Section, this
tool allows us to straightforwardly find a new meron solution generalizing the spherically symmetric Gauss--Bonnet vacuum and a recently found five-dimensional Einstein meron. 

\subsection{Extended Thermodynamics} \label{secc:ET}

In this Section, we turn to the study of the meronic black hole solution using
the Euclidean quantum gravity approach. By Wick rotating the black hole we find that
the Euclidean time period must be
\begin{equation}
 \beta=\frac{4\pi l^2 r_+^3}{3r_+^4+l^2(r_+^2-\rho^2)}. \label{beta}
\end{equation}
This quantity corresponds to the inverse temperature of the black hole, $T=\beta^{-1}$. In other words, the temperature is calculated \`a la Hawking $T=f'(r_+)/4\pi$.
With this in mind, we calculate the Euclidean
action in accordance to Section \ref{secc:ABC}, considering the bulk action, the surface terms and the AdS counterterms
\begin{equation}
 I_E=I_B+I_{ST}+I_{CT}.
\end{equation}
For the YM meron at hand we obtain
\begin{equation}
 I_E=\frac{\beta}{2G}\left[Gm-\frac{r_+^3}{l^2}+\frac{\rho^2}{r_+}
 -\alpha\left( \frac{8G^2m^2}{r_+^3} -\frac{12Gm\rho^2}{r_+^4} +\frac{4\rho^4}{r_+^5}
 -\frac{4r_+^3}{l^4}+\frac{4Gm}{l^2} \right)\right], \label{4GBaction}
\end{equation}
and after some algebraic manipulations
\begin{equation}
 I_E=\beta m-\frac{\pi}{G}\left(r_+^2+4\alpha \right).
\end{equation}
It comes to no surprise that the characteristic energy of the system is the black hole mass, i.e.,
\begin{equation} \label{enthalpy}
 H\equiv\left(\frac{\partial I_E}{\partial\beta}\right)_{p,\alpha}=m.
\end{equation}
Even in four dimensions, we know that the GB parameter serves as a modification to the
Bekenstein area formula~\cite{Castro:2013pqa}. In deed, the entropy for the meronic black hole
is
\begin{equation} \label{entropy}
 S\equiv\beta\left(\frac{\partial I_E}{\partial\beta}\right)_{p,\alpha}-I_E
 =\frac{\pi}{G}\left(r_+^2+4\alpha \right).
\end{equation}
This result coincides with the Wald entropy.
The thermodynamic volume of the system is
\begin{equation} \label{volume}
  V\equiv\frac{1}{\beta}\left(\frac{\partial I_E}{\partial p}\right)_{\beta,\alpha}
  =\frac{4\pi r_+^3}{3}.
\end{equation}
This variable is unaffected by the addition of $\alpha$ into the fold, contrasting with the entropy. The final equation of state is for the dual of $\alpha$
\begin{equation} \label{dualofalpha}
  \psi\equiv-\frac{1}{\beta}\left(\frac{\partial I_E}{\partial\alpha}\right)_{\beta,p}
  =\frac{4\pi T}{G},
\end{equation}
which is consistent with equation \eqref{psi} of the Appendix. Notice that $\psi$
is completely determined by the temperature. This is similar to how radiation pressure in
a photon gas is fixed exclusively by the temperature. The previous equation signals a
decrease in thermodynamic degrees of freedom. For example, the GB parameter is completely determined once $S$ and $V$ are fixed. Alternatively, the thermodynamic volume is totally determined for a given pair $S,\alpha$.

Proceeding as in reference~\cite{Smarr:1972kt}, determines the following Smarr relation
\begin{equation}
 \frac{H}{2}=\frac{\kappa A}{8\pi G}-pV+E, \label{class-smarr}
\end{equation}
where $\kappa$ is the surface gravity of the black hole and $A$ the horizon area. 
We have also used equation \eqref{YMenergy} as the meron's energy appears above. 
Nonetheless, equations \eqref{enthalpy}-\eqref{dualofalpha} allow for this equation
to be written in Gibbs--Duhem form
\begin{equation}
 \frac{H}{2}=TS-pV-\alpha\psi+E. \label{smarrGB4D}
\end{equation}
The equations of state are consistent with the association of the Euclidean action with
the Gibbs free energy of the system, i.e., ${\cal G}=I_E/\beta=H-TS$ and with the following
first law of thermodynamics
\begin{equation}
 \rd H=T\rd S+V\rd p-\psi\rd\alpha. \label{firstlawGB4D}
\end{equation}

\begin{figure}[ht]
\begin{minipage}[c]{0.42\linewidth}
\includegraphics[scale=0.5, trim={0 0 0 0.5cm},clip]{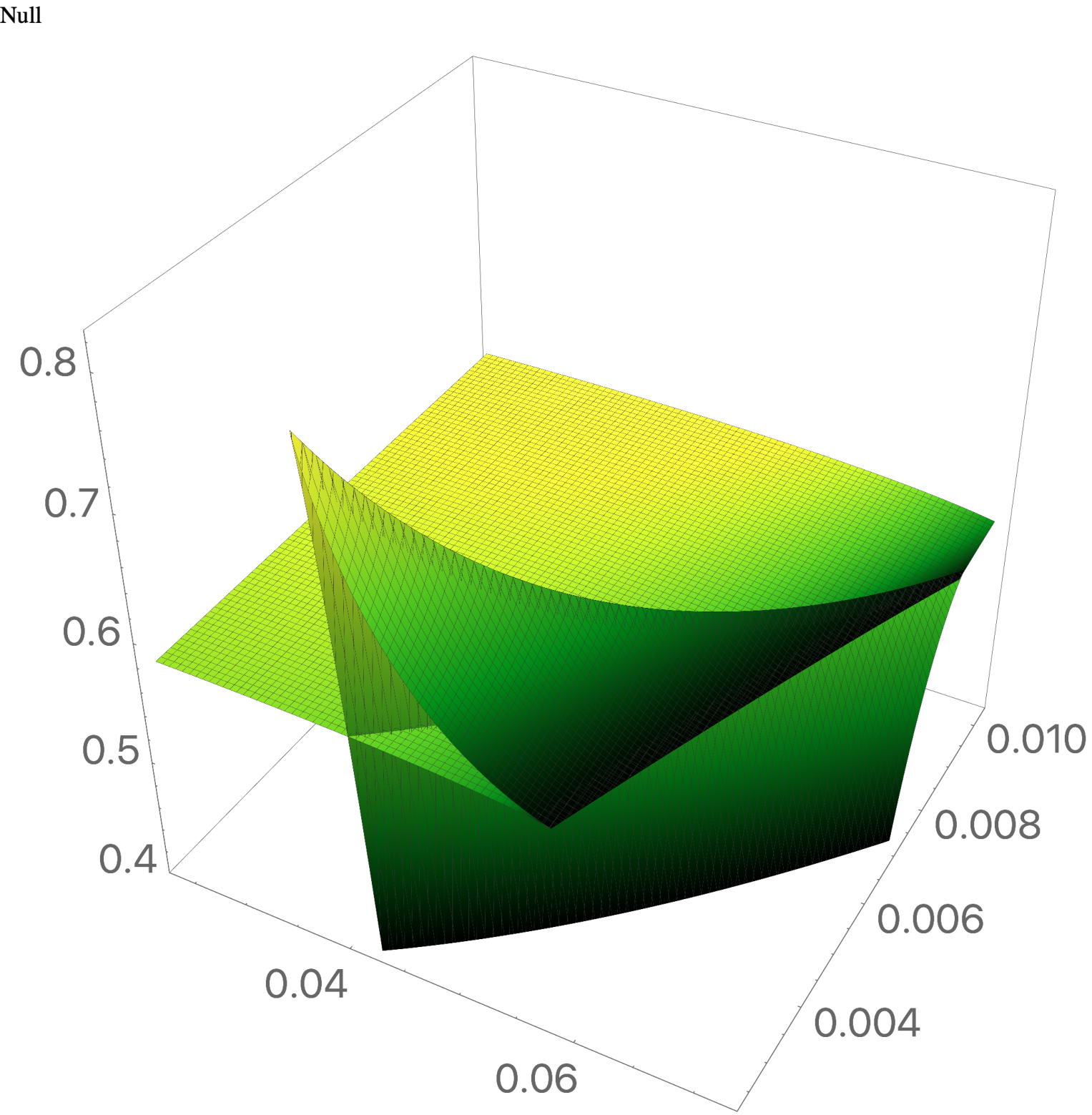}
\put(-225,130){$\cal G$}
\put(-135,10){$T$}
\put(-20,45){$p$}
\end{minipage}
\hskip1cm
\begin{minipage}[c]{0.43\linewidth}
\includegraphics[scale=0.5, trim={0 0 0 0.5cm},clip]{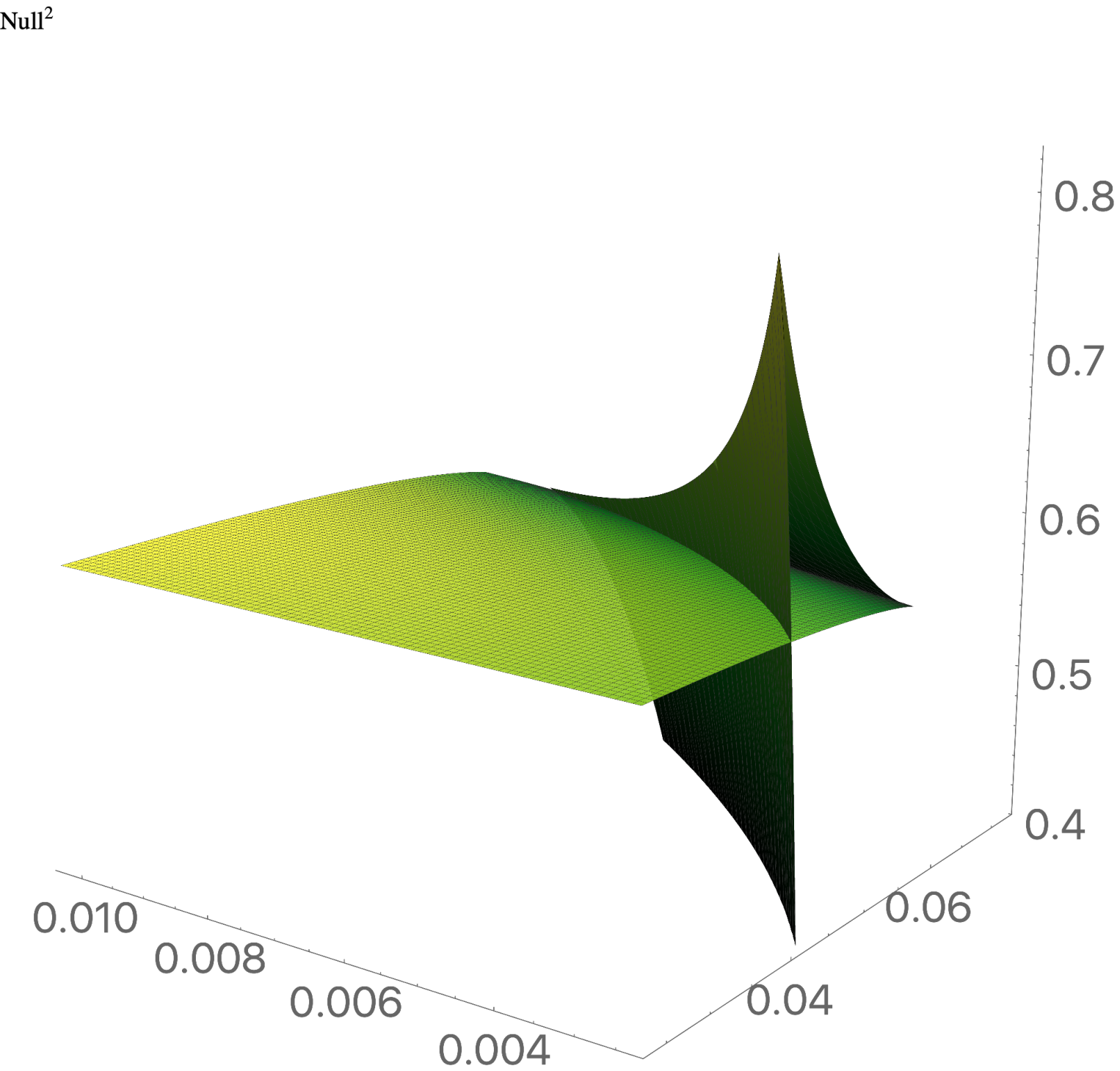}
\put(-10,160){$\cal G$}
\put(-165,10){$T$}
\put(-45,30){$p$}
\end{minipage}%
\caption{Free energy $\cal G$ against temperature $T$ and pressure $p$ are displayed for the meron in Gauss--Bonnet theory with Reissner--Nordstr\"om geometry. Here we have used $G = 1$, $\rho =0.578$ and $\alpha=-0.1$.
The left panel allows for a more detailed visualization of the swallowtail cross sections. The right panel offers a view of the characteristic curve formed by the Hawking--Page-like transitions.}
\label{fig:4Dmeron3Dplot}
\end{figure}

In the sequel, we illustrate how the Gibbs free energy behaves near criticality.
First off, notice that equation \eqref{beta} determines the horizon radius as a function
the thermodynamic variables. The equation is a quartic polynomial in $r_+$ yet one of the solutions is always
negative. Since this is mathematically unsensible three characteristic sizes exist for
the meronic black hole. The usual parlance is to dub the three branches: the small, intermediate and large black holes. 
In figure \ref{fig:4Dmeron3Dplot} we plot the Gibbs free energy of the system.
There it is shown that 
the intermediate black hole always has greater
free energy than the other two. Thus, it is under no condition thermodynamically prefered.
For low temperatures the small black hole dominates and for high temperatures it is the large one.
There is a first order phase transition of Hawking--Page-type~\cite{Hawking:1982dh} between small and large black holes.

%%%%%%%%%%%%%
%%%%%%%%%%%%%

\begin{figure}[ht]
\begin{minipage}[c]{0.35\linewidth}
\includegraphics%[width=\linewidth, trim={1cm 0 0 0},clip]{BH_zero_alpha.eps}
[scale=0.45,trim={0 0 0 0.5cm},clip]{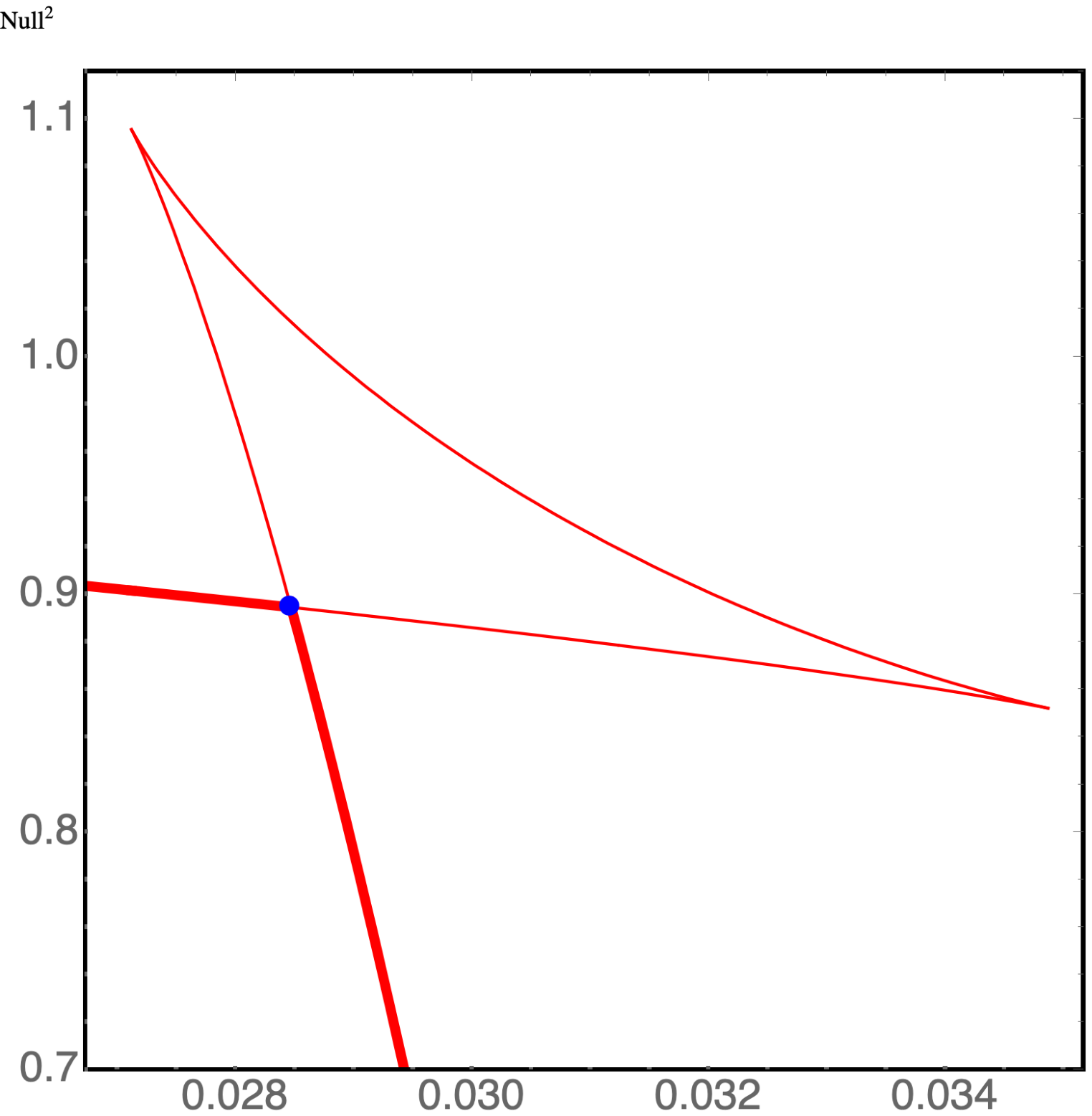}
\put(-173,100){\tiny SBH}
\put(-133,55){\tiny LBH}
\put(-100,-10){$T$}
\put(-200,165){$\cal G$}
\end{minipage}
\hskip1cm
\begin{minipage}[c]{0.35\linewidth}
\includegraphics%[width=\linewidth, trim={1cm 0 0 0},clip]{BH_phase.eps}
[scale=0.4615,trim={0 0 0 0.5cm},clip]{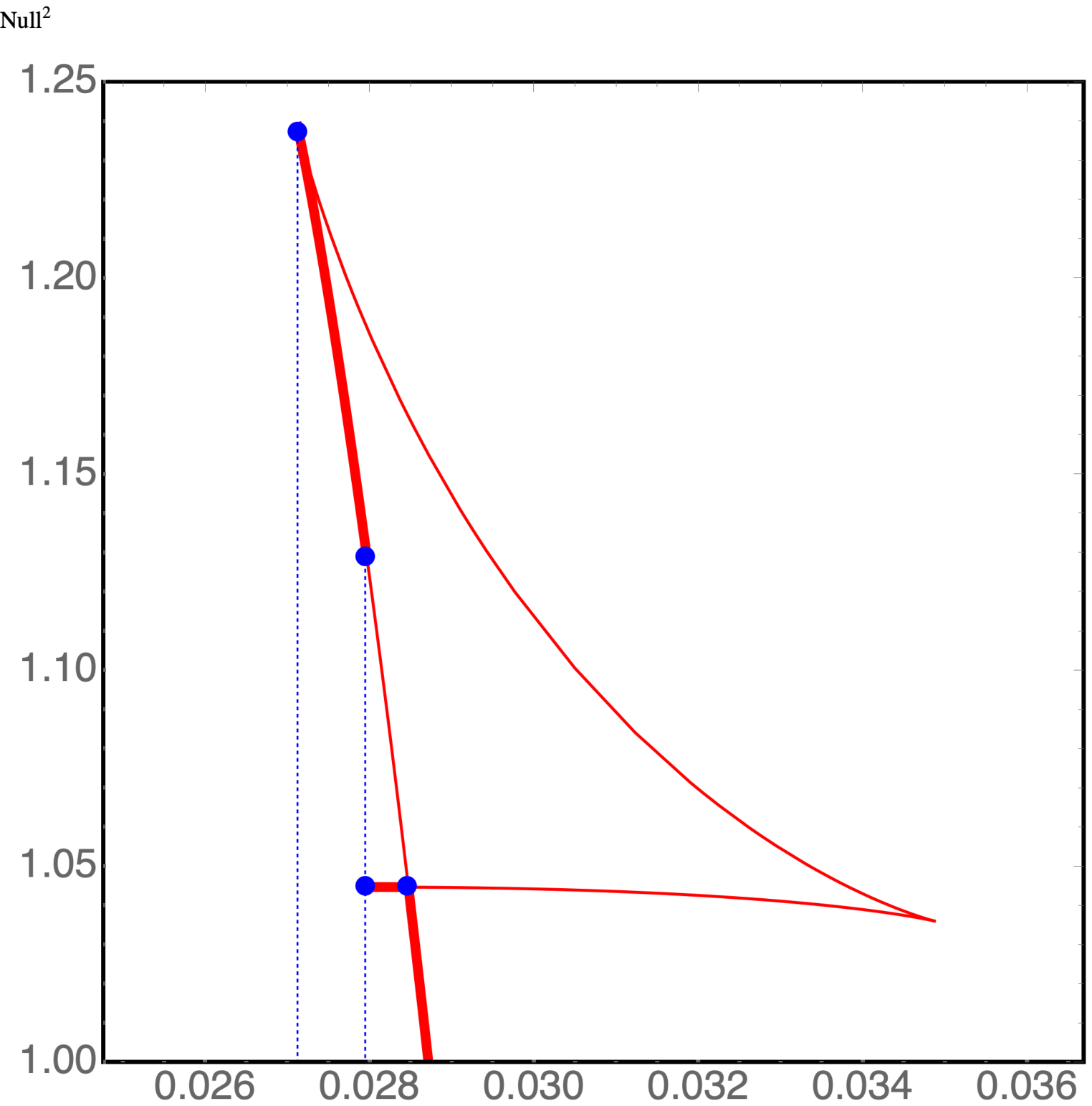}
\put(-100,-10){$T$}
\put(-200,170){$\cal G$}
\end{minipage}%
\caption{Plots of free energy vs. temperature with $G=1$, $\rho=1$, $l=10$. The left panel shows a typical small black hole (SBH) to large black hole (LBH) phase transition when $\alpha=0$. The right panel
shows a reentrant phase transition of the type LBH/SBH/LBH when $\alpha=-0.42$.}
\label{fig:4Dmeron2Dplot}
\end{figure}

In the left panel of figure \ref{fig:4Dmeron2Dplot} the small/large phase transition is plotted
for a fixed pressure but varying temperature. A different type of phase transition
occurs in the system if we
allow for negative values of $\alpha$ and restrict the entropy to be positive~\cite{Xu:2019xif}.
This restriction conduces to a reentrant phase transition
which we depict in the right panel of figure \ref{fig:4Dmeron2Dplot}.

\section{The Five-Dimensional Black Hole} \label{secc:5D}

In the previous Section, we presented a meron adapted to the nested spheres of a static spacetime.
We saw that the YM meron contributes through an inverse square term in the metric function.
This allowed the matter content to mirror a Dirac monopole.
We continue with this strategy here, choosing a meron well adapted to a spherically symmetric background. However, as we see below, the meron backreacts in the metric with a term that goes
as $\ln{r}/r^2$. In five dimensions, this can no longer be interpreted as electromagnetic matter.
In fact, this is closer to a mass term which goes as an inverse square. Thus, the meron's selfgravity
requires a careful definition of gravitational charge.

We use Schwarzschild coordinates \eqref{SchCoords} for the black hole geometry 
\begin{equation}
 g=-f(r)\rd t\otimes \rd t+f(r)^{-1}\rd r\otimes \rd r+r^2\gamma_{3},
\end{equation}
and choose Euler angles $(\Psi,\Theta,\Phi)$ to chart $\gamma_3$. The gauge potential has the form
$A=\lambda U^{-1}\rd U$ as in \eqref{omegaU} with an SU(2)-valued scalar field
\begin{equation}
U=\exp\left(\Phi\, \rt^3\right)\exp\left(\Theta\, \rt^2\right)\exp\left(\Psi\, \rt^3\right).
\end{equation}
In other words, the meron is proportional to the left-invariant Maurer--Cartan form of SU(2).
In comparison with the meron of Section \ref{secc:4D}, this configuration employs a generalization of the hedgehog Ansatz~\cite{Ayon-Beato:2015eca,Canfora:2017yio}. This configuration presents a \emph{gravitational spin from isospin}
effect\cite{Canfora:2018ppu} where fermionic excitation arises as a bound state of two bosons: a scalar and a meron.
This phenomenon is only possible when the meron self-gravitates. From a holographic point of view this configuration is
also worth examining. The spin from isospin effect in the bulk may lead to the computation
of fermionic observables of the boundary CFT using a purely bosonic theory.

The YM equation of motion \eqref{YMeom} fixes $\lambda=1/2$ while
the gravitational field equation \eqref{EOM} determines the following Wheeler polynomial
\begin{equation}
 \frac{1}{l^2}+{\cal F}+\ta {\cal F}^2=\frac{s}{r^4}+\frac{2\rho^2\ln{r}}{r^4}. \label{GBWP}
\end{equation}
The roots of this polynomial yield the following metric functions
\begin{equation}
 f(r)=1+\frac{r^2}{2\ta}\left(1\pm\sqrt{1-\frac{4\ta}{l^2}+\frac{4\ta [s+2\rho^2\ln{r}]}{r^4}}\right),
\end{equation}
and from now on we only consider the branch with the negative sign.
This is often the most studied branch solution as it connects continuously with
Einstein theory. Like all spherically symmetric solutions of GB theory, the metric
function has the form
\begin{equation}
 f(r)=1+\frac{r^2}{2\ta}-\frac{1}{2\ta}\sqrt{Q(r)}.
\end{equation}
Thus, the black hole has two curvature singularities: the one at $r=0$ and another
at $r=r_c$, where $Q(r_c)=0$. For the present meronic black hole $r_c$ is always
hidden behind the event horizon when $\alpha$ is positive. This is the prescribed phenomenology from string theory. However, if only for generality, when $\alpha$
is negative the singularity will remain covered by the horizon if $\alpha<-L^2/8$.

Now, drawing inspiration from reference~\cite{Canfora:2018ppu} we reparameterize $s$
so that the metric function takes the form
\begin{equation}
 f(r)=1+\frac{r^2}{2\ta}\left(1-\sqrt{1-\frac{4\ta}{l^2}+\frac{4\ta [8mG/3\pi
 +(1+L^2/l^2)L^2+2\rho^2\ln{r/L}]}{r^4}}\right). \label{GBmeron}
\end{equation}
Following the last reference exactly would yield a similar result and is obtained
by a shift in $s$ by $2\alpha$.
In equation \eqref{GBmeron} $L$ is defined by
\begin{equation}\label{defL}
 L^2=\frac{l}{4}\left(-l+\sqrt{l^2+8\rho^2}\right),
\end{equation}
the reason is given further below.
Now, notice that when $\rho\to0$ then $L\to0$ which is useful for evaluating \eqref{GBmeron}
in this limit, i.e.,
\begin{equation}
 f(r)\to1+\frac{r^2}{2\ta}\left(1-\sqrt{1-\frac{4\ta}{l^2}+\frac{4\ta [8mG/3\pi]}{r^4}}\right).
\end{equation}
This is the branch of the Boulware--Deser black hole which smoothly connects to
the Schwarzschild--Tangherlini solution when $\ta\to0$. 
In this equation, our parametrization shows $m$ to be the black hole mass. Now, by taking the Einstein limit of \eqref{GBmeron} we obtain
\begin{equation}\label{Emeron}
 f(r)\to1-\frac{8Gm}{3\pi r^2}-\frac{(1+L^2/l^2)L^2+2\rho^2\ln{r/L}}{r^2}+\frac{r^2}{l^2},
\end{equation}
thus recuperating the geometry of the meronic black hole in~\cite{Canfora:2018ppu}.
This is to say, equation \eqref{GBmeron} is the interpolation between these two spherically symmetric black holes.

The two merons we present in this manuscript are complementary in the following way.
The YM energy of the four-dimensional meron is finite but its winding number is trivial. The meron in this Section has winding number 1 and its Chern number is 1/2 which make it interesting from a topological point of view. However, it is less manageable than the one in Section \ref{secc:4D} as its energy is infinite. This in turn make the mass of the black hole infinite. Nonetheless, sensible thermodynamics can be extracted from this type of solution.
The key is to consider the black hole of the smallest size (the least entropy) as a reference geometry.
This extremal black hole occurs at zero temperature. The enthalpy of formation of any finite
temperature black hole coincides with the mass difference between the two black holes.
In equation \eqref{Emeron} $m$ represents precisely this amount, the difference between
the black hole mass and that of the zero temperature black hole. This approach was taken in~\cite{Emparan:1999pm,Flores-Alfonso:2019aae} and could have been taken in the previous Section
but there the distinction was not crucial due to finite YM energy. In what follows we
provide further detail on this matter.

The temperature of the black hole is fixed by the period of the imaginary time circle
in the Euclidean sheet of the solution. This period is determined by demanding that the gravitational instanton be free of conical singularities in the Euclidean version of the
event horizon, thus
\begin{equation} \label{GBT}
 T=\frac{f'(r_+)}{4\pi}=\frac{2r_+^4+l^2(r_+^2-\rho^2)}{2\pi l^2 r_+(r_+^2+4\alpha)}.
\end{equation}
Notice that the temperature and the horizon radius are related through a quartic polynomial.
Restricting the radius to be positive determines that the minimum size of the black hole
occurs at zero temperature. We designate this extremal value by $L$ and \eqref{defL} is now
justified. The mass difference between a black hole of size $r_+$ and one of size $L$ is
\begin{equation}
 \Delta M=m-\frac{3\pi\alpha}{4G}. \label{deltam}
\end{equation}
The parametrization of~\cite{Canfora:2018ppu} is such that $\Delta M$ is dubbed $m$ in the
metric function. As mentioned above, this is achieved in \eqref{GBmeron} shifting the integration constant $s$ by $2\alpha$. This mass difference is motivated from a thermodynamical point of view as in the extended framework mass corresponds to enthalpy of formation. Pictorially, one may think of black hole mass as the amount of energy it takes
to cut away a region of spacetime to form a black hole\cite{Kastor:2009wy,Johnson:2014xza}. In this scenario it is natural to find the spacetime with no
horizon or the smallest one. However, there is an additional motivation from this one which
comes from field theory. In spite of the infinite YM energy each configuration possesses, the energy 
difference between any two configurations is finite. Harmonizing these two ideas results in us calculating
the YM energy difference
\begin{equation} \label{deltae}
 \Delta E=-\frac{3\pi\rho^2}{4G}\ln{\left(\frac{r_+}{L}\right)}.
\end{equation}

\subsection{More on the Thermodynamics} \label{secc:Mott}

Before continuing with the thermodynamics of the Gauss--Bonnet meron, let us consider its Einstein limit, given by equation \eqref{Emeron}. The methods of Section \ref{secc:ABC} yield a finite Euclidean gravitational action but the matter action diverges. However,
subtracting the on shell action of the extremal black hole yields a finite result,
\begin{equation}
 \frac{\Delta I_E}{\beta}=\frac{1}{2G}\left[\frac{2Gm}{3}-\frac{\pi^2(r_+^4-L^4)}{2l^2}
 -\pi\rho^2\ln{r_+/L}\right]=m-TS.
\end{equation}
The main purpose the subtraction serves is to yield a finite thermodynamic energy, a sensible notion of enthalpy of formation $H\equiv\partial\Delta I_E/\partial \beta=m$. The volume is defined as in \eqref{volume}
by $\Delta V\equiv\partial\Delta I_E/\partial p$. We also adopt the notation
$\Delta V=V-V_e$ to highlight the role of the extremal black hole solution. Hence,
we write
\begin{equation} \label{deltav}
 V-V_e=\frac{\pi^2\left(r_+^4-L^4\right)}{2}.
\end{equation}

Returning to equation \eqref{GBmeron} we may proceed as in~\cite{Cvetic:2001bk} given that our black hole
is five dimensional. However, an alternative regularization is given by a Kounterterm series~\cite{Kofinas:2006hr}, boundary terms which contain extrinsic geometric information. In general, the advantage is that
there is a universal form for the boundary terms for any given dimension and any Lovelock theory~\cite{Kofinas:2008ub}. In odd dimensions, this method gives rise to characteristic vacuum energies with thermodynamical
implications~\cite{Mora:2004rx}. However, since we perform an action subtraction this energy is not in play here.
Thus, we write the action difference as
\begin{equation}\label{GBYMaction}
 \Delta I_E= \beta\left(m-\frac{3\pi\alpha}{4G}\right)-\frac{\pi^2r_+}{2G}\left(r_+^2+12\alpha\right),
\end{equation}
yielding the enthalpy of formation, i.e., as
\begin{equation} \label{GBH}
 H=\frac{3\pi(r_+^4-L^4)}{8Gl^2}+\frac{3\pi(r_+^2-L^2)}{8G}
 -\frac{3\pi\rho^2\ln{r_+/L}}{4G}.
\end{equation}
Moreover, the entropy of the system can be directly read off equation \eqref{GBYMaction} as
\begin{equation} \label{defS}
 S=\frac{\pi^2r_+}{2G}\left(r_+^2+12\alpha\right).
\end{equation}
This equation possesses the same form as the
Boulware--Deser entropy. This is also true of its charged version, the Wiltshire
solution~\cite{Wiltshire:1985us}, which was 
studied from the perspective of the standard references~\cite{Cai:1998vy,Cai:2001dz} in~\cite{Johnson:2015ekr}.
The idea behind this procedure is also used in Appendix \ref{secc:Boulware--Deser} to derive an expression for the thermodynamic dual of $\alpha$, which for the present
solution means
\begin{equation}
 \Delta\psi=\psi-\psi_e=\frac{6\pi^2r_+}{G}T.
\end{equation}
To sum up, the first law of thermodynamics is
\begin{equation}
 \rd H=T\rd S+(V-V_e)\rd p-(\psi-\psi_e)\rd\alpha. \label{firstlaw}
\end{equation}
This equation is structurally comparable to the first law of thermodynamics coming from
the fixed charge ensemble of AdS black holes~\cite{Chamblin:1999tk}. With these thermodynamic quantities in mind we write down its Smarr relation as
\begin{equation}
 H=\frac{3}{2}TS-p\Delta V-\alpha\Delta\psi+\Delta E, \label{smarr}
\end{equation}
where we have used \eqref{deltae}. Notice that, as in Section \ref{secc:4D}, the meron's simple nature manifests itself through almost hair-like thermodynamic equations. This is to say, that
the YM matter does not show itself in the first law [cf. \eqref{firstlawGB4D} and \eqref{firstlaw}], so at a glance, it appears to be a black hole hair.
However the merons' Smarr relations \eqref{smarrGB4D} and \eqref{smarr} show this
is a misconception.

\begin{figure}[ht]
\centering
\includegraphics[scale=0.5,trim={0 0 0 0.5cm},clip]{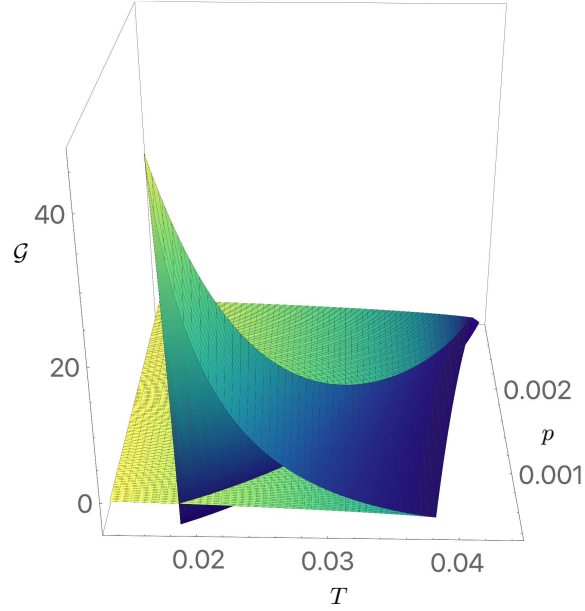}
\put(-20,52){$p$}
\put(-100,-10){$T$}
\put(-220,120){$\cal G$}
\caption{The Gibbs free energy of our meronic Gauss--Bonnet solution exhibiting swallowtail behaviour in the subcritical regime. The thermodynamic potential $\cal G$ is plotted against temperature $T$ and pressure $p$. The intersection of planes representing large and small black holes is a curve designating Hawking--Page transitions, here we chose the values $G=1$, $\rho=0.5$ and $\alpha=1$.}
\label{fig:5Dmeron3Dplot}
\end{figure}

\begin{figure}[ht]       
    \fbox{\includegraphics[width=.3\textwidth, trim={0.25cm 0 0 0.35cm},clip]{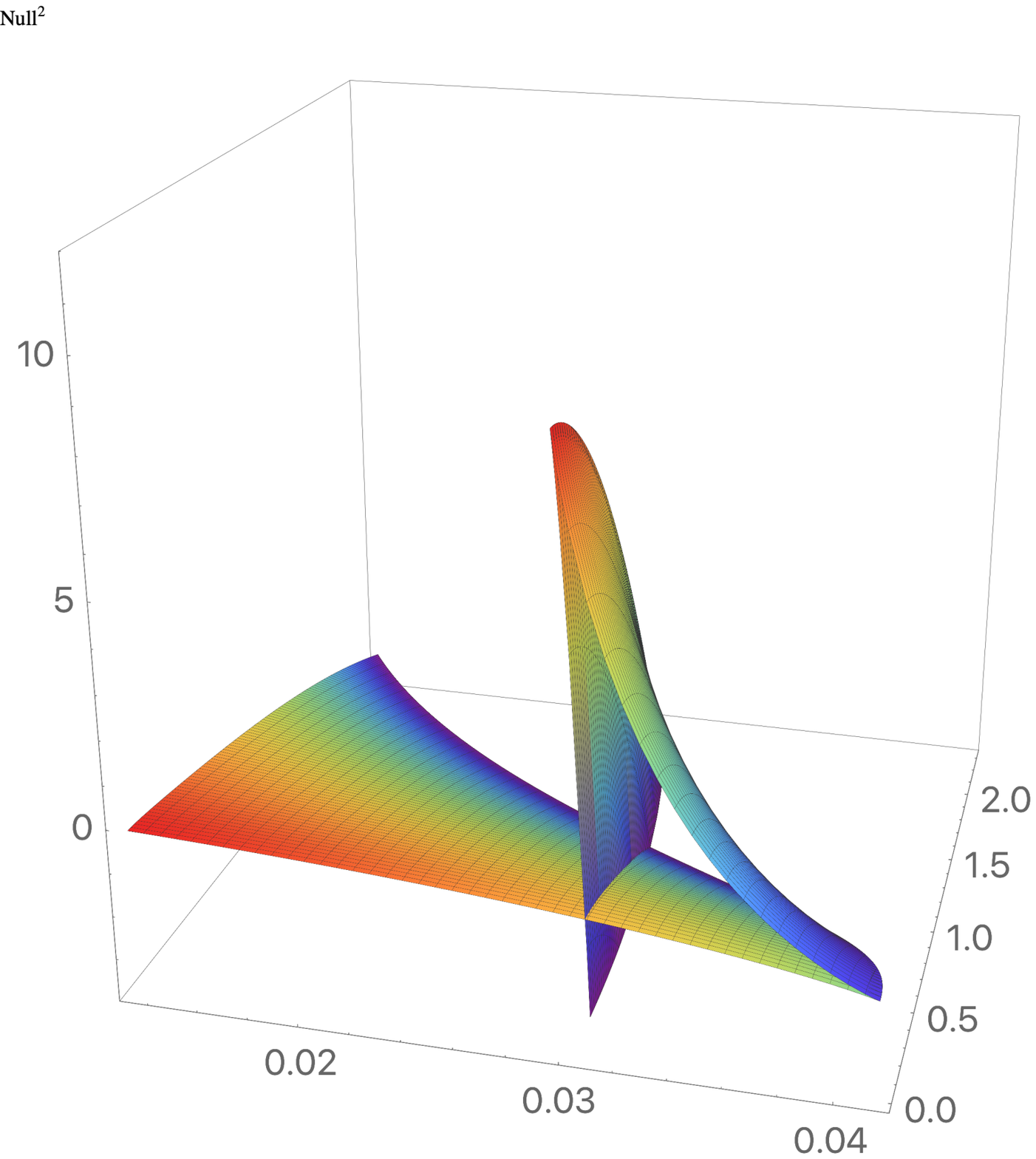}}   
    \put(-7,42){$\rho$}
\put(-100,0){$T$}
\put(-170,120){$\cal G$}
    \hspace{5px}
    \fbox{\includegraphics[width=.3\textwidth, trim={0.25cm 0 0 0.35cm},clip]{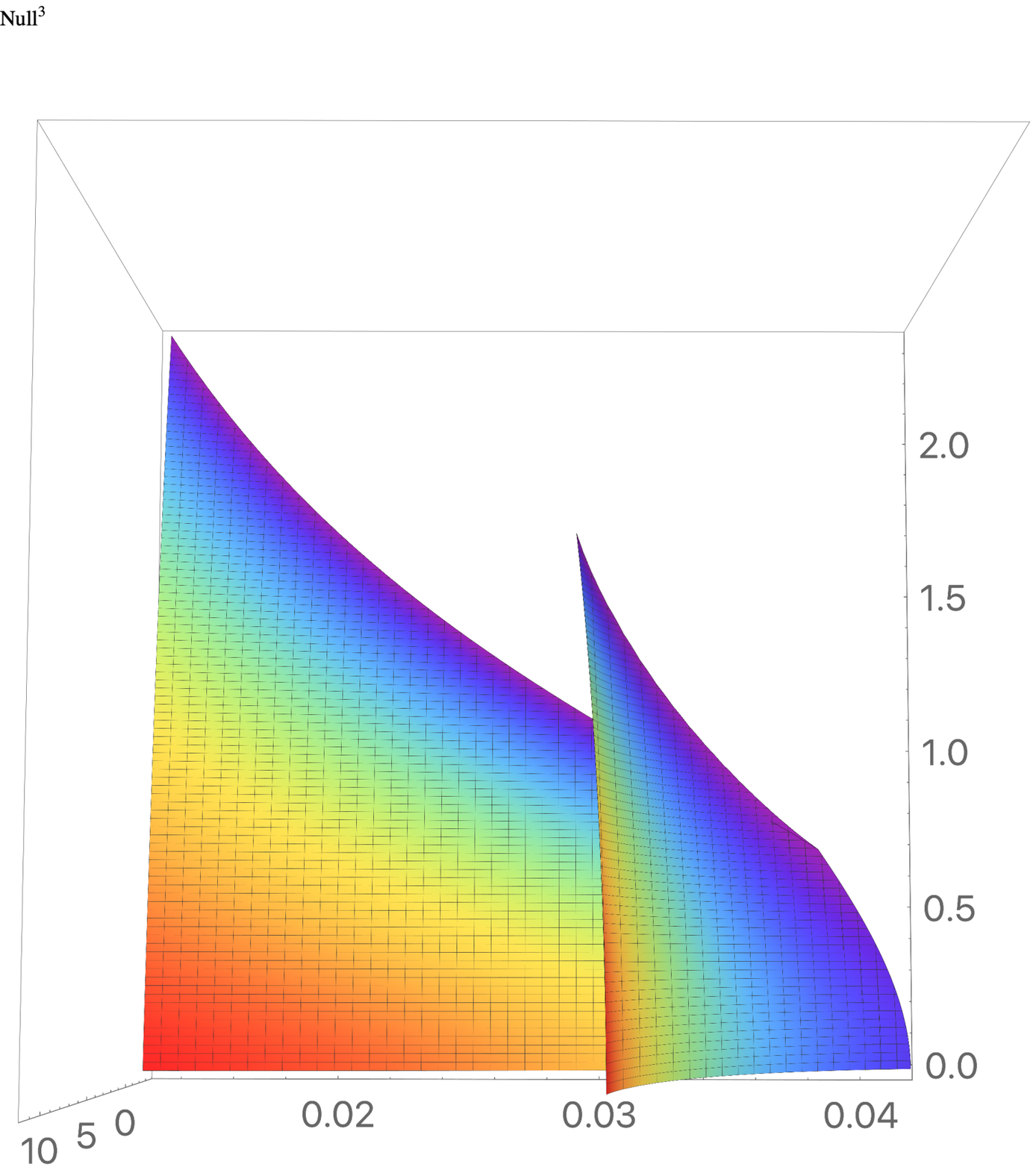}}
    \put(-20,80){$\rho$}
\put(-90,0){$T$}
\put(-165,10){$\cal G$}
    \hspace{5px}
    \fbox{\includegraphics[width=.3\textwidth, trim={0.25cm 0 0 0.35cm},clip]{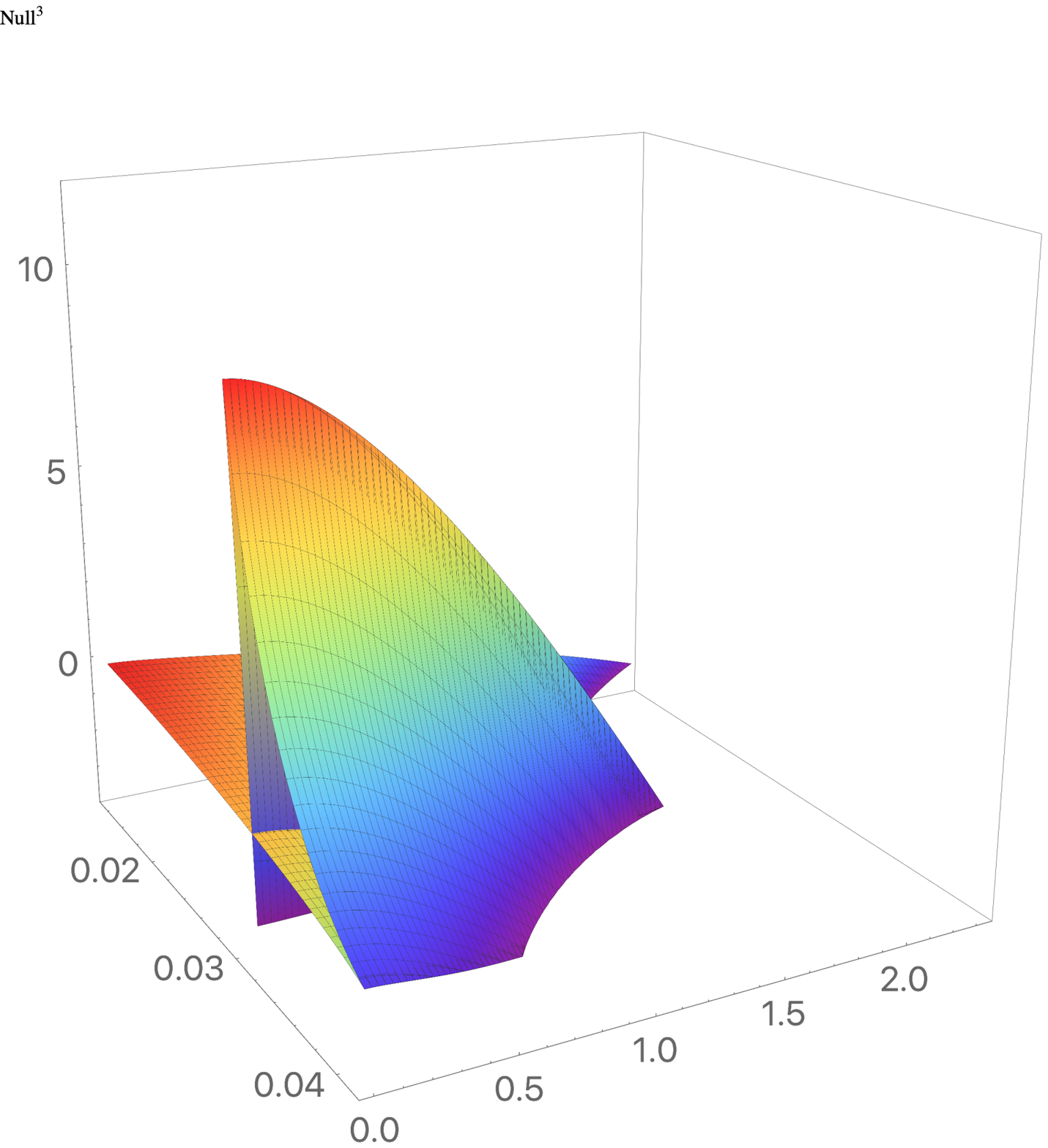}}
    \put(-50,22){$\rho$}
\put(-150,30){$T$}
\put(-160,130){$\cal G$}
    \caption{Multiple views of the Gibbs free energy $\cal G$ plotted against temperature $T$ and the coupling parameter $\rho=G/2e^2$ with $G=1$, $\rho=0.5$, $\alpha=1$ and $p=0.001$. The left panel shows the characteristic swallowtail behaviour of the thermodynamic potential. The middle panel offers a ``bird's eye view'' of the figure showing how as the coupling parameter grows larger the swallowtail structure disappears. The right panel displays how the swallowtail
    shrinks leaving a continuous curve beyond a critical value of $\rho$.}
    \label{fig:5Dmeron3Dplot_rho}
\end{figure}

The Gibbs free energy ${\cal G}=H-TS$ of this system is plotted in figure \ref{fig:5Dmeron3Dplot} where it shows
typical swallowtail behaviour in the subcritical regime. In a complementary manner, we display in figure \ref{fig:5Dmeron3Dplot_rho} how the free energy conducts itself from subcritical to supercritical values of $\rho$.

%%%%%%%%%%%%%%
%%%%%%%%%%%%%%

%%%%%%%%%%%%%
%%%%%%%%%%%%

\subsection{Special Couplings} \label{SpclCpls}

To finalize this section, we offer a word on very particular values of the Lovelock coupling constants.
In five dimensions, the Gauss--Bonnet action \eqref{LLaction} corresponds to
Lanczos--Lovelock theory with coupling constants $\alpha_0=-2\Lambda$, $\alpha_1=1$
and $\alpha_2=\alpha$. The Lovelock parameters couple the cosmological constant, the Einstein--Hilbert and the Gauss--Bonnet terms together. The special class $\alpha_1=0$
is known as pure Lovelock gravity~\cite{Kastor:2006vw,Cai:2006pq}. Where we have allowed for a possibly non-zero cosmological constant.

The Wheeler polynomial \eqref{Wheeler} is given by
\begin{equation} \label{WPL}
 \frac{1}{l^2}+2\alpha{\cal F}^2=\frac{8mG/3\pi
 +(1+L^2/l^2)L^2+2\rho^2\ln{r/L}}{r^4},
\end{equation}
where we have chosen to parameterize the integration constant as in equation \eqref{GBmeron}. 
From it the temperature of the pure Lovelock (PL) black hole is determined to be
\begin{equation} \label{PLT}
 T_{\rm{PL}}=\frac{2r_+^4-\rho^2l^2}{8\pi l^2\alpha r_+}.
\end{equation}
This implies $L$ in equation \eqref{WPL} is given by
$L^4=\rho^2l^2/2$ so that it corresponds to the
horizon radius of the extremal black hole. Basing ourselves in Section \ref{secc:Mott} we find the enthalpy of the system to be
the black hole mass difference from the extremal configuration. Thus, we write
\begin{equation}
 H_{\rm{PL}}=-\frac{3\pi\rho^2\ln{r_+/L}}{4G}+\frac{3\pi(r_+^4-L^4)}{8Gl^2}.
\end{equation}
The entropy of the black hole is analogous to \eqref{defS}
\begin{equation}
 S_{\rm{PL}}=\frac{6\pi^2\alpha r_+}{G}.
\end{equation}
Notice the absence of the Bekenstein area contribution which comes from the Einstein dynamics.
Aside from the equations just above and the observation that $\psi_e=0$ every other equation of state from 
Section \ref{secc:Mott} remains the same. In deed, the first law of thermodynamics and the Smarr relation
for this black hole read as equations \eqref{firstlaw} and \eqref{smarr}, respectively. Nonetheless, a key difference
in the thermodynamics is the phase structure. Equation \eqref{PLT} is a quartic equation for $r_+$, the discriminant
of this relation is negative meaning only two real solutions exist to the equation. However, only one of these solutions
is positive. In other words, the system is single-phased.
%%%%%%%%%%%%%%%%%%%%%%%%%%%%%%%%%%%%%%%%%%%%%%%%%%%%%%%%%%%%%%%%%%%
%%%%%%%%%%%%%%%%%                          %%%%%%%%%%%%%%%%%%%%%%%%
%%%%%%%%%%%%%%%%% modify from here onwards %%%%%%%%%%%%%%%%%%%%%%%%
%%%%%%%%%%%%%%%%%                          %%%%%%%%%%%%%%%%%%%%%%%%
%%%%%%%%%%%%%%%%%%%%%%%%%%%%%%%%%%%%%%%%%%%%%%%%%%%%%%%%%%%%%%%%%%%

Another specialization of the Lanczos--Lovelock functional \eqref{LLaction} is the gravitational Chern--Simons action, which
only exists in odd spacetime dimensions~\cite{Garraffo:2008hu}. The characteristic coupling constants are given by $\alpha_0=-2\Lambda$, $\alpha_1=1$ and $\alpha_2=-3/4\Lambda$. 
It is very well-known that the relation among the couplings arises from requiring the theory to have the maximum possible number of degrees of freedom~\cite{Troncoso:1999pk}.
Notice that Einstein theory cannot be recovered from this action, e.g., neither the limit $\Lambda\to0$ nor $\Lambda\to\infty$ are realizable. We can think of this Chern--Simons gravity as a five-dimensional analogue of the Chern--Simons (CS) description of three-dimensional Einstein theory~\cite{Witten:1988hc}. For further details on this theory we refer the reader to~\cite{Zanelli:2005sa}.

The Wheeler polynomial for the CS class is
\begin{equation} \label{WCS}
 \frac{1}{l^2}\left(1+\frac{l^2{\cal F}}{2}\right)^2=\frac{8mG/3\pi
 +(1+L^2/l^2)L^2+2\rho^2\ln{r/L}}{r^4},
\end{equation}
allowing us to write down the Hawking temperature as
\begin{equation}
 T_{\rm{CS}}=\frac{2r_+^4+l^2(r_+^2-\rho^2)}{\pi l^2 r_+(2r_+^2+l^2)},
\end{equation}
which is just equation \eqref{GBT} evaluated at the CS condition, i.e., $\alpha=l^2/8$.
Notice that once again, the temperature is related to the black hole horizon radius through a quartic
function. The discriminant of this quartic is always negative determining that the CS black hole
has only one phase. Now, the enthalpy of formation $H$ is given by
the black hole mass difference from the extremal case.
The equation just above determines $L$ to be just as in Section \ref{secc:Mott} and $H$
is given by \eqref{GBH}. Hence, the entropy is
\begin{equation}
 S_{\rm{CS}}=\frac{\pi^2r_+}{2G}\left(r_+^2+\frac{3l^2}{2}\right),
\end{equation}
just equation \eqref{defS} evaluated at the CS condition, as expected.

In contrast, the thermodynamic volume suffers the greatest departure from its value in the general Gauss--Bonnet case.
Entropy and pressure are not independent variables, so we expect some contribution from entropy
to be reflected in the volume, indeed we have
\begin{equation}
 \Delta V_{\rm{CS}}= \frac{\pi^2\left(r_+^4-L^4\right)}{2}+T_{\rm{CS}}\left(\pi^3l^4r_+\right).
\end{equation}
This complies with the first law of thermodynamics in the form
\begin{equation}
 \rd H=T\rd S+(V-V_e)\rd p. \label{firstlawCS}
\end{equation}
and is also consistent with the Gibbs--Duhem equation
\begin{equation}
 H=\frac{3}{2}TS-p\Delta V+\Delta E. \quad [\rm{for~CS}]
\end{equation}
We mention that 
following a similar process to the one found in Appendix \ref{secc:Boulware--Deser}
one finds
\begin{equation}
 \Delta V= \left(\frac{\partial H}{\partial p}\right)_{r_+}-T\left(\frac{\partial S}{\partial p}\right)_{r_+}. 
 \quad [\rm{for~CS}]
\end{equation}
When the cosmological constant is fixed we recover the first law in the form $\rd H=T\rd S$
with the absence of YM energy contributions as is characteristic of merons. In the limit $\rho\to0$ we obtain the vacuum case, see for example reference~\cite{Crisostomo:2000bb}.
It is noteworthy that even though the CS theory has a higher (gauge) symmetry
than GB it does not lead to more global charges. Thus, as we have seen above, there is no
additional charge entering the thermodynamics.

As a final comment, we mention that in this work we have focused on the metric formulation (torsion-free) of Gauss--Bonnet theory. However, one may consider a more general situation in which torsion is present. Finding (stable) solutions with torsion is in general more complicated. Our present solution bears many geometric features with the half-BPS solutions found in~\cite{Canfora:2007xs}. Thus, an extension of our present solution is desirable
along these lines. To minimally couple torsion to the Yang--Mills sector could be problematic from the gauge symmetry point of view. To preserve the gauge symmetry one possibility is to have the torsion uncoupled from the YM matter. That said, the equations of motion for the connection do not imply that the matter content sources the torsion. Thus, within the firs order formalism our solution may be recovered with zero torsion. Further still an extension
with a non-Abelian gauge field and torsion is (in principle) obtainable. It would be
interesting to see how the presence of torsion modifies the present black hole.

\section{Conclusions} \label{secc:Conc}
In this work, we study spherically symmetric black holes with SU(2) Yang--Mills matter
in Einstein--Gauss--Bonnet theory.
The configurations are meronic implying they are simple in nature yet intrinsically
non-Abelian. The first black hole we examine is the four-dimensional Einstein meron
which we inject into Gauss--Bonnet theory. The principal effect is that the black hole entropy is modified by the theory's parameter $\alpha$. However, this modification
is consistent with the classic thermodynamic equations. The framework of extended thermodynamics supplements this by providing the necessary conceptual structure for a new, consistent,
interpretation of the first law of thermodynamics and the system's Gibbs--Duhem equation.
Under the assumption that the black hole entropy must remain positive always the regime where $\alpha<0$ leads to a reentrant phase transition. For high temperatures large black holes dominate the path integral. Lowering the temperature conducts to a Hawking--Page transition into a small black hole. However, further lowering the temperature leads to a reentrance into the large black hole phase until the systems arrives to its lowest possible temperature.

In the second part of this investigation, we generalize a recently found five-dimensional
Einstein meron to a Gauss--Bonnet version. Turning off the matter content yields the Boulware--Deser solution. As a Yang--Mills field the meron is topologically non-trivial yet it possesses infinite energy. 
However, the Euclidean quantum gravity approach is able to deal with this difficulty.
The way this meron backreacts
on spacetime closely resembles the way mass usually contributes to the metric function. Ultimately, the infinite YM energy is the source of difficulty within the thermodynamics. However, a sensible notion of mass is given by considering the extremal black hole of the configuration. This extremal black hole
does not give off Hawking radiation. The mass difference between any black hole and the extremal provides a consistent notion of mass and enthalpy. Enthalpy is derived from the Euclidean action subtraction
and coincides with the mass difference. 
As far as the thermodynamics is concerned, our solution exhibits the critical behaviour of van der Waals fluids and charged AdS black holes. This result further strengthens
the statement that black holes lie within the van der Waals universality class.

In reference~\cite{Flores-Alfonso:2019aae} compatibility was found between the Euclidean and the Lorentzian quasilocal~\cite{Kim:2013zha,Gim:2014nba,Hyun:2017nkb} approaches. A treatment of meronic black holes a long the lines of quasilocal methods is desirable. We mention that for planar black holes the Smarr relation in the presence of a cosmological constant takes the form
\begin{equation}
 (D-1)M=(D-2)TS, \label{class-Smarr-D}
\end{equation}
which is thermodynamically consistent with the first law of black hole mechanics in the classical framework. A generalized version of the previous equation for Lifshitz 
black holes has been recently found~\cite{Ayon-Beato:2019kmz}. Within the extended framework
the equivalent form of the same Smarr relation is found in~\cite{Hyun:2017nkb}. The pair of equations we reference are comparable, e.g., to our equations \eqref{class-smarr} and \eqref{smarrGB4D}. We emphasize, for clarity, that equation \eqref{class-Smarr-D} is satisfied by the BTZ black hole~\cite{Banados:1992wn}. A shortcoming of our approach in this manuscript
is that although the cosmological constant can be understood as a constant of motion, as discussed above, it is not clear if this is applicable for the Gauss--Bonnet sector.

During the preparation of this work a recent paper was published~\cite{Ayon-Beato:2019tvu}.
Therein, the authors construct SU(3) self-gravitating Skyrmions.
A comparison is carried out between trivially embedded SU(2) solutions into SU(3) and nonembedded solutions. These configurations are closely related to meronic Yang--Mills black holes. We shall further explore this topic in upcoming investigations.

\section*{Acknowledgments}
We are in debt with Eloy Ay\'on-Beato, Fabrizio Canfora, Crist\'obal Corral and Hernando Quevedo for interesting comments and helpful discussions.
DFA would like to thank the Mexican Secretariat of Public Education (Secretar\'ia de Educaci\'on P\'ublica) for support under PRODEP project No. 12313509.
He is also grateful to the Centro de Estudios Cient\'ificos, for its hospitality during the completion of this work.

\appendix

\section{Extended Thermodynamics of Asymptotically Flat Boulware-Deser Black Holes}
\label{secc:Boulware--Deser}

The mass of asymptotically flat Schwarzschild--Tangherlini black holes is completely determined by the horizon geometry, $M=M(r_+)$.
This is also true for their entropies, $S=S(r_+)$, hence mass is a function only of entropy $M=M(S)$ which yields
\begin{equation}
 \rd M=\frac{\rd M}{\rd S}\rd S,
\end{equation}
ultimately yielding
\begin{equation}
 T\equiv\frac{\rd M}{\rd S}.
\end{equation}

The Boulware--Deser family generalizes these black holes to a spherically symmetric, and so static, class of spacetimes
which are now additionally parameterized by the Gauss--Bonnet (GB) parameter $\alpha$. Treating this physical parameter as a thermodynamic
variable modifies the previous equations in that now $M=M(r_+,\alpha)$ and $S=S(r_+,\alpha)$. In general, mass is a function
of both entropy $S$ and the GB coupling, $\alpha$. Thermodynamics is said to be \emph{extended} by this consideration. The first law of thermodynamics
is generalized to
\begin{equation}
 \rd M=T\rd S -\psi\rd \alpha,
\end{equation}
where $T$ is the temperature and $\psi$ is the thermodynamic conjugate of $\alpha$.

Given the new functional dependence among $M$, $S$ and $\alpha$ let us combine the following equations
\begin{equation}
 \rd M=\left(\frac{\partial M}{\partial r_+}\right)_{\alpha}\rd r_+ 
 +\left(\frac{\partial M}{\partial \alpha}\right)_{r_+}\rd \alpha,
\end{equation}
and
\begin{equation}
 \rd S=\left(\frac{\partial S}{\partial r_+}\right)_{\alpha}\rd r_+ 
 +\left(\frac{\partial S}{\partial \alpha}\right)_{r_+}\rd \alpha.
\end{equation}
This leads to the expression
\begin{equation} \label{diffM}
 \rd M=\left(\frac{\partial M}{\partial r_+}\right)_{\alpha} 
 \left(\frac{\partial S}{\partial r_+}\right)^{-1}_{\alpha}
 \left[ \rd S-\left(\frac{\partial S}{\partial \alpha}\right)_{r_+}\rd\alpha \right]
 +\left(\frac{\partial M}{\partial \alpha}\right)_{r_+}\rd \alpha.
\end{equation}
In this context, temperature is given by
\begin{equation}
 T\equiv \left(\frac{\partial M}{\partial S}\right)_{\alpha} 
 =\left(\frac{\partial M}{\partial r_+}\right)_{\alpha} 
 \left(\frac{\partial S}{\partial r_+}\right)^{-1}_{\alpha},
\end{equation}
so that equation \eqref{diffM} becomes
\begin{equation}
 \rd M=T\rd S - \left[ T\left(\frac{\partial S}{\partial \alpha}\right)_{r_+} -
 \left(\frac{\partial M}{\partial \alpha}\right)_{r_+}  \right]\rd \alpha.
\end{equation}
Lastly, we have that
\begin{equation}
 \psi\equiv \left(\frac{\partial M}{\partial \alpha}\right)_{S} 
 = T\left(\frac{\partial S}{\partial \alpha}\right)_{r_+} - 
 \left(\frac{\partial M}{\partial \alpha}\right)_{r_+}. \label{psi}
\end{equation}

\bibliography{GaussBonnet}

\end{document}